\documentclass[twocolumn]{aastex631}
\usepackage{threeparttable}
\usepackage{newtxtext,newtxmath}
\usepackage[T1]{fontenc}
\usepackage{nicefrac}
\usepackage{latexsym}
\usepackage{amsmath, mathrsfs, bm}
\usepackage{eqnarray}
\usepackage{url}
\usepackage{cases}
\usepackage{epsfig}
\usepackage{color}
\usepackage{graphicx}
\usepackage{grffile}
\usepackage{float}
\usepackage{soul}
\usepackage{enumerate}
\usepackage{multirow}
\usepackage{xspace}
\usepackage{bm}
\usepackage{times}

\usepackage{etoolbox}
\usepackage{multirow}

\begin{document}

\title{Prospects for Probing the Interaction between Dark Energy and Dark Matter Using Gravitational-wave Dark Sirens with Neutron Star Tidal Deformation}

\author{Tian-Nuo Li}
\affiliation{Key Laboratory of Cosmology and Astrophysics (Liaoning) \& College of Sciences, Northeastern University, Shenyang 110819, People's Republic of China}

\author{Shang-Jie Jin}
\affiliation{Key Laboratory of Cosmology and Astrophysics (Liaoning) \& College of Sciences, Northeastern University, Shenyang 110819, People's Republic of China}

\author{Hai-Li Li}
\affiliation{Basic Teaching Department, Shenyang Institude of Engineering, Shenyang 110136, People's Republic of China}

\author{Jing-Fei Zhang}
\affiliation{Key Laboratory of Cosmology and Astrophysics (Liaoning) \& College of Sciences, Northeastern University, Shenyang 110819, People's Republic of China}

\author{Xin Zhang}
\affiliation{Key Laboratory of Cosmology and Astrophysics (Liaoning) \& College of Sciences, Northeastern University, Shenyang 110819, People's Republic of China}
\affiliation{Key Laboratory of Data Analytics and Optimization for Smart Industry (Ministry of Education), Northeastern University, Shenyang 110819, People's Republic of China}
\affiliation{National Frontiers Science Center for Industrial Intelligence and Systems Optimization, Northeastern University, Shenyang 110819, People's Republic of China}

\correspondingauthor{Jing-Fei Zhang}
\email{jfzhang@mail.neu.edu.cn}

\correspondingauthor{Xin Zhang}
\email{zhangxin@mail.neu.edu.cn}

\begin{abstract}
Gravitational wave (GW) standard siren observations provide a rather useful tool to explore the evolution of the Universe. In this work, we wish to investigate whether dark sirens with neutron star (NS) deformation from third-generation GW detectors could help probe the interaction between dark energy and dark matter. We simulate the GW dark sirens of four detection strategies based on 3 yr observation and consider four phenomenological interacting dark energy (IDE) models to perform cosmological analysis. We find that GW dark sirens could provide tight constraints on $\Omega_{\rm m}$ and $H_0$ in the four IDE models, but do not perform well in constraining the dimensionless coupling parameter $\beta$ in models of the interaction proportional to the energy density of cold dark matter. Nevertheless, the parameter degeneracy orientations of cosmic microwave background (CMB) and GW are almost orthogonal, and thus, the combination of them could effectively break cosmological parameter degeneracies, with the constraint errors of $\beta$ being $0.00068-0.018$. In addition, we choose three typical equations of state (EoSs) of an NS, i.e., SLy, MPA1, and MS1, to investigate the effect of an NS's EoS in cosmological analysis. The stiffer EoS could give tighter constraints than the softer EoS. Nonetheless, the combination of CMB and GW dark sirens (using different EoSs of an NS) shows basically the same constraint results of cosmological parameters. We conclude that the dark sirens from 3G GW detectors would play a crucial role in helping probe the interaction between dark energy and dark matter, and the CMB+GW results are basically not affected by the EoS of an NS.

\end{abstract}

\section{Introduction}\label{sec:intro}
The precise measurements of the cosmic microwave background (CMB) anisotropies have ushered in the era of precision cosmology \citep{WMAP:2003ivt,WMAP:2003elm}. However, recent precise measurements of the cosmological parameters lead to some puzzling tensions. Especially for the measurement of the Hubble constant, the values of the Hubble constant inferred from the Planck CMB observations (assuming the $\Lambda$CDM model; \citealt{Planck:2018vyg}) and obtained through the distance ladder measurements (model-independent; \citealt{Riess:2021jrx}) are in more than $5\sigma$ tension \citep{Riess:2021jrx}, which is now commonly considered a severe crisis for cosmology \citep{Riess:2019qba,Verde:2019ivm}; see e.g., \citet{Yang:2018euj}, \citet{Guo:2018ans,Guo:2019dui}, \citet{Riess:2019qba}, \citet{Verde:2019ivm}, \citet{cai:2020}, \citet{Ding:2019mmw}, \citet{DiValentino:2019jae,DiValentino:2019ffd,DiValentino:2021izs}, \citet{Feng:2019jqa}, \citet{Hryczuk:2020jhi}, \citet{Li:2020tds}, \citet{Lin:2020jcb}, \citet{Liu:2019awo}, \citet{Vagnozzi:2019ezj,Vagnozzi:2021gjh}, \citet{Zhang:2019cww}, \citet{Cai:2021wgv}, \citet{Gao:2021xnk}, \citet{Abdalla:2022yfr}, \citet{Vagnozzi:2021tjv}, \citet{Wang:2021kxc}, and \citet{Zhao:2022yiv} for recent related discussions. On the other hand, the cosmological constant $\Lambda$ in the $\Lambda$CDM model is equivalent to the vacuum energy also sufferred from serious theoretical challenges, namely, the ``fine-tuning'' and ``cosmic coincidence'' problems \citep{Sahni:1999gb,Bean:2005ru}. Therefore, it is far-fetched to take the $\Lambda$CDM model as the eventual cosmological scenario. All of these facts imply that new physics beyond the $\Lambda$CDM model needs to be considered. Meanwhile, some promising cosmological probes also need to be further developed to provide independent measurements of the Hubble constant.

Among various extensions to the $\Lambda$CDM model, there is a class of models known as the interaction dark energy (IDE) models, in which the direct and nongravitational interaction between dark energy and dark matter is considered.
On the one hand, the IDE models can not only help alleviate the Hubble tension \citep{Guo:2018gyo,Feng:2019mym} but also alleviate the coincidence problem of dark energy \citep{Cai:2004dk,Zhang:2005rg}. On the other hand, the investigation of the IDE models can help probe the fundamental nature of dark energy and dark matter. Although CMB can give rather tight constraints on the $\Lambda$CDM model, the newly extended parameters beyond the $\Lambda$CDM model will severely degenerate with other parameters when the CMB data alone are used to constrain the extended cosmological models. Hence, other cosmological probes need to be combined with the CMB data to break the cosmological parameter degeneracies.

As a promising cosmological probe, gravitational-wave (GW) standard sirens will play a crucial role in exploring the expansion history of the Universe; see e.g., \citet{Dalal:2006qt}, \citet{Cutler:2009qv}, \citet{Zhao:2010sz,Zhao:2018gwk,Zhao:2019gyk}, \citet{Cai:2016sby,Cai:2017plb}, \citet{Cai:2017aea,Cai:2018rzd}, \citet{Chang:2019xcb}, \citet{Du:2018tia}, \citet{He:2019dhl}, \citet{Yang:2019vni,Yang:2019bpr}, \citet{Zhang:2019ylr}, \citet{Bachega:2019fki}, \citet{Borhanian:2020vyr}, \citet{Chen:2020dyt}, \citet{Gray:2019ksv}, \citet{Jin:2020hmc,Jin:2021pcv,Jin:2022tdf,Jin:2022qnj,Jin:2023zhi,Jin:2023tou,Jin:2023sfc}, \citet{Chen:2020zoq}, \citet{Mitra:2020vzq}, \citet{Qi:2021iic}, \citet{Ye:2021klk}, \citet{Cao:2021zpf}, \citet{Colgain:2022xcz}, \citet{deSouza:2021xtg}, \citet{Dhani:2022ulg}, \citet{Wang:2021srv,Wang:2022oou,Wang:2022rvf}, \citet{Zhu:2021bpp}, \citet{Califano:2022syd}, \citet{Han:2023exn}, \citet{Hou:2022rvk}, 
 \citet{Wu:2022dgy}, \citet{Zhang:2023gye}, and \citet{Song:2022siz} for related discussions. The luminosity distance can be directly obtained from the analysis of the GW waveform without having to rely on the distance ladder, hence the name, standard siren \citep{Schutz:1986gp,Holz:2005df}. The redshift information of the GW source can be precisely measured from the observations of the electromagnetic (EM) counterparts (in this case, the event is usually referred to as a bright siren), and thus the multimessenger observation can enable us to establish the distance-redshift relation to constrain cosmological parameters. The first multimessenger event GW170817, together with its EM counterparts \citep{LIGOScientific:2017vwq,LIGOScientific:2017ync}, has given an independent measurement on the Hubble constant, $H_0=70^{+12}_{-8}~\rm km~s^{-1}~Mpc^{-1}$ with a precision of 14\% \citep{LIGOScientific:2017adf}. If 50 similar bright sirens are present, the constraint precision of the Hubble constant can be improved to 2\% \citep{Chen:2017rfc}. However, it is not sufficient to use the current standard sirens to probe the interaction between dark energy and dark matter, and one has to resort to the next-generation GW detectors.

In the next decades, the third-generation (3G) detectors, the Einstein Telescope (ET; \citealt{Punturo:2010zz}) and the Cosmic Explorer (CE; \citealt{LIGOScientific:2016wof}) will be operational, with sensitivities improved one order of magnitude over the current GW detectors and a large number of GW events can be observed \citep{Evans:2021gyd}. Nevertheless, the detection of EM counterparts is rather difficult. Recent forecasts show that only $\sim 0.1\%$ of binary neutron star (BNS) mergers detected by the 3G detectors can have detectable EM counterparts \citep{Yu:2021nvx}. Moreover, bright sirens can only give a tight constraint on the Hubble constant but do not perform well in constraining other cosmological parameters.

In general, only BNS and black hole (BH)-NS merger events have the potential for EM counterparts, and in most cases, these events are not observable in EM waves. Binary black hole (BBH) merger events do not have EM counterparts (it should be noted that the events of supermassive or massive BH binaries may have observable EM counterparts; see, e.g., \citealt{Tamanini:2016zlh}, \citealt{Wang:2019tto,Wang:2021srv,Wang:2022oou}, and \citealt{Zhao:2019gyk}). For GW events without observable EM counterparts, there are actually methods to conduct cosmological studies using standard sirens. The commonly employed approach is to statistically determine the redshifts of GW events by cross-correlating them with galaxy catalogs; see, e.g., \citet{LIGOScientific:2018gmd}, \citet{DES:2019ccw}, \citet{DES:2020nay}, \citet{LIGOScientific:2019zcs,KAGRA:2021duu}, and \citet{Palmese:2021mjm}. This type of GW standard siren is referred to as a dark siren.

In fact, it is possible to establish the distance-redshift relation solely through GW observations without the need for EM observations, including galaxy catalogs. \citet{Messenger:2011gi} proposed that in BNS merger events, the additional tidal phase caused by the tidal deformation of an NS can aid in determining the redshifts of GW events. Consequently, both luminosity distance and redshift can be independently determined from GW observations alone. Since this method does not involve EM counterpart observations, in this paper, we also refer to such standard sirens as dark sirens. It is important to note that the dark sirens discussed in this paper are distinct from the dark sirens using the galaxy catalog method.

Therefore, employing this method, all the detected BNS mergers can be used to perform cosmological analysis; see, e.g., \citet{DelPozzo:2015bna}, \citet{Wang:2020xwn}, \citet{Chatterjee:2021xrm}, \citet{Dhani:2022ulg} \citet{Ghosh:2022muc}, and \citet{Jin:2022qnj} for recent works. However, the tidal deformation is related to the equation of state (EoS) of NS, which may lead to different constraint results. By solely using the dark sirens, the EoS of dark energy and the Hubble constant can be precisely measured, provided that the EoS of an NS is perfectly known \citep{Jin:2022qnj}.

These dark sirens show a strong ability to constrain cosmological parameters. Therefore, in this work, we wish to investigate the potential of the dark sirens in constraining the IDE models. The aim of this work is to answer three key questions: (i) what precisions the cosmological parameters in the IDE models could be measured by solely using the dark sirens, (ii) how the GW dark sirens break the cosmological parameter degeneracies generated by the EM observations, and (iii) how the EoS of an NS affects the constraints on the cosmological parameters. Through this work, we wish to shed light on these aspects and explore the potential role of the dark sirens in probing the interaction between dark energy and dark matter.

This work is organized as follows. In Sec.~\ref{sec2}, we introduce the methodology used in this work. In Sec.~\ref{sec3}, we report the constraint results and make some relevant discussions. The conclusion is given in Sec.~\ref{sec4}.

\section{Methodology and data}\label{sec2}
\subsection{Interacting Dark Energy Models}
In a flat Friedmann--Roberston--Walker universe, the Friedmann equation is given by
\begin{equation}\label{2.1}
3M^2_{\rm{pl}} H^2=\rho_{\rm{de}}+\rho_{\rm c}+\rho_{\rm b}+\rho_{\rm r},
\end{equation}
where $3M^2_{\rm{pl}} H^2$ is the critical density of the universe, and $\rho_{\rm{de}}$, $\rho_{\rm c}$, $\rho_{\rm b}$, and $\rho_{\rm r}$ represent the energy densities of dark energy, cold dark matter, baryon, and radiation, respectively.

\begin{figure*}[htbp]
\includegraphics[scale=0.85]{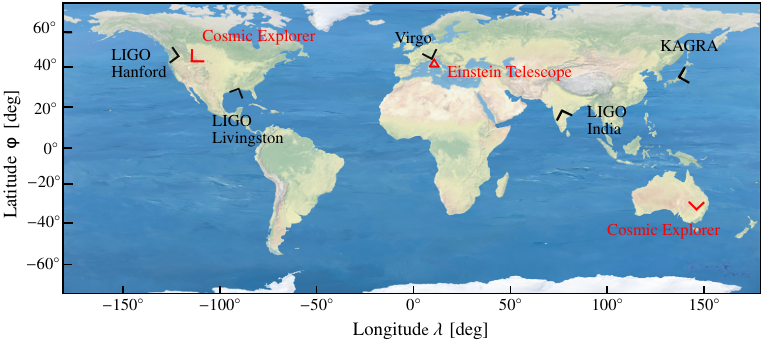}
\centering
\caption{\label{fig1} Locations and orientations of the ground-based GW observatories. The second-generation GW detectors are shown in black lines \citep{lalsuite,Ashton:2018jfp} and the 3G GW detectors considered in this work are shown in red lines \citep{di2021seismological,Borhanian:2020ypi}.}
%The locations and orientations for LIGO Hanford, LIGO Livingston, Virgo, and KAGRA are taken from Ref.~\citep{lalsuite}, for LIGO India are taken from Ref.~\citep{Ashton:2018jfp}, for ET are taken from Ref.~\citep{di2021seismological}, and for CE are taken from Ref.~\citep{Borhanian:2020ypi}. The detailed coordinates of the GW observatories are listed in Table \ref{tab1}.
\end{figure*}

In the IDE models, if assuming a direct interaction between dark energy and cold dark matter, the energy conservation equations can be given by
\begin{align}\label{conservation1}
\dot{\rho}_{\rm de} +3H(1+w)\rho_{\rm de}= Q,\\
\dot{\rho}_{\rm c} +3 H \rho_{\rm c}=-Q,
\end{align}
where the dot is the derivative with respect to the cosmic time $t$, $H$ is the Hubble parameter, $w$ is the EoS of dark energy, and $Q$ is the energy transfer rate. In this work, we consider four typical phenomenological forms of $Q$, i.e., $Q=\beta H\rho_{\rm de}$ (I$\Lambda$CDM1), $Q=\beta H\rho_{\rm c}$ (I$\Lambda$CDM2), $Q=\beta H_0\rho_{\rm de}$ (I$\Lambda$CDM3), and $Q=\beta H_0\rho_{\rm c}$ (I$\Lambda$CDM4). Here $\beta$ is a dimensionless coupling parameter that describes the interaction strength between dark energy and dark matter. $\beta > 0$ indicates cold dark matter decaying into dark energy, $\beta < 0$ indicates dark energy decaying into cold dark matter, and $\beta = 0$ indicates no interaction between dark energy and cold dark matter. Note that we wish to investigate the potential of GW dark sirens in probing the interaction (or constraining $\beta$), and thus we consider $w=-1$ for simplicity.

In the IDE models, there is a problem of early-time perturbation instability \citep{Majerotto:2009zz,Clemson:2011an} because the cosmological perturbations of dark energy in the IDE models will be divergent in a part of the parameter space, which leads to the ruin of the IDE cosmology in the perturbation level. To avoid this problem, \citet{Li:2014eha,Li:2015vla,Li:2023fdk} extended the parameterized post-Friedmann (PPF) approach \citep{Fang:2008sn,Hu:2008zd} to the IDE models, referred to as the ePPF approach. This approach can safely calculate the cosmological perturbations in the whole parameter space of the IDE models. In this work, we employ the ePPF approach to treat the cosmological perturbations; see e.g., \citet{Zhang:2017ize,Feng:2018yew} for the application of the ePPF approach.

%\begin{figure*}[htbp]
\begin{figure}[]
\includegraphics[scale=0.45]{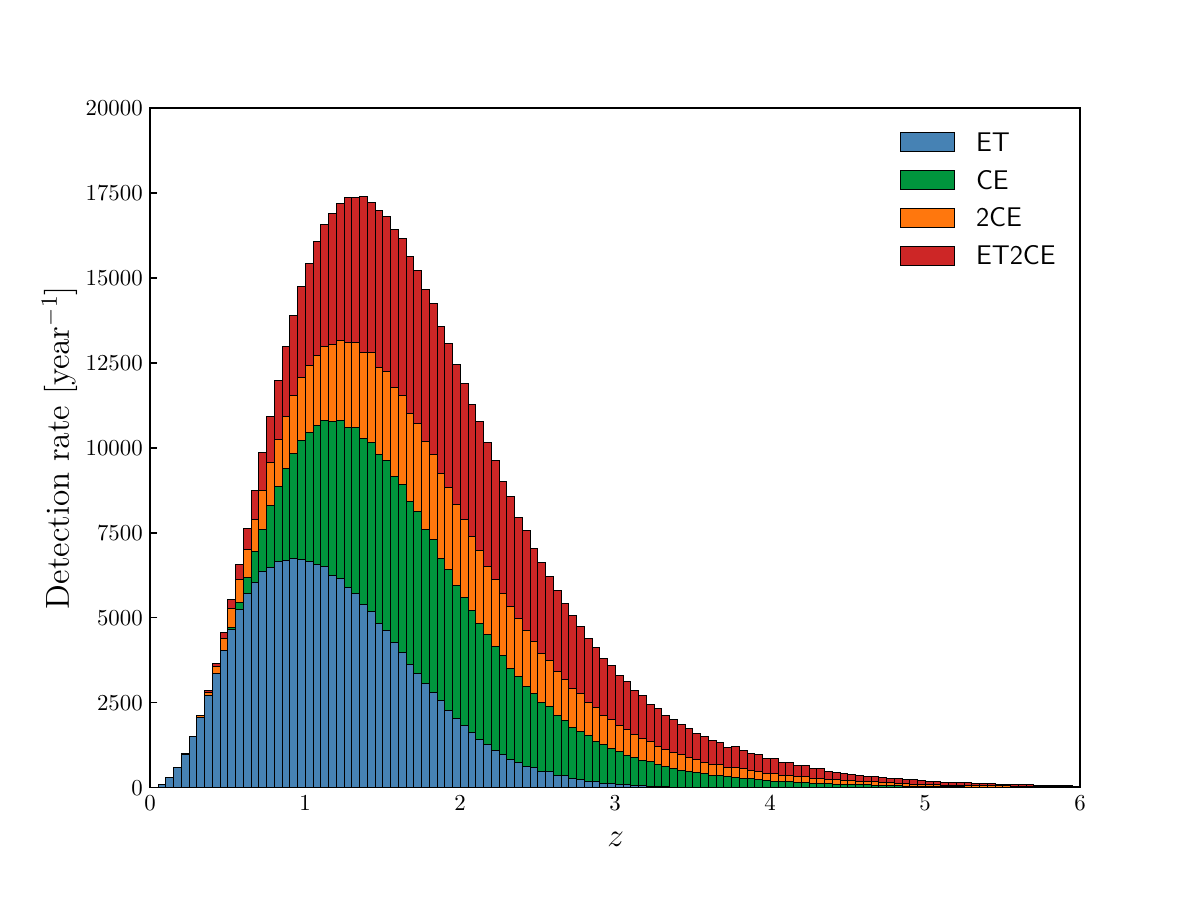}
\centering
\caption{\label{fig2} Expected detection rates as a function of redshift for ET, CE, 2CE, and ET2CE per year.}
\end{figure}

\begin{figure}
\includegraphics[width=0.51\textwidth,height=6.5cm]{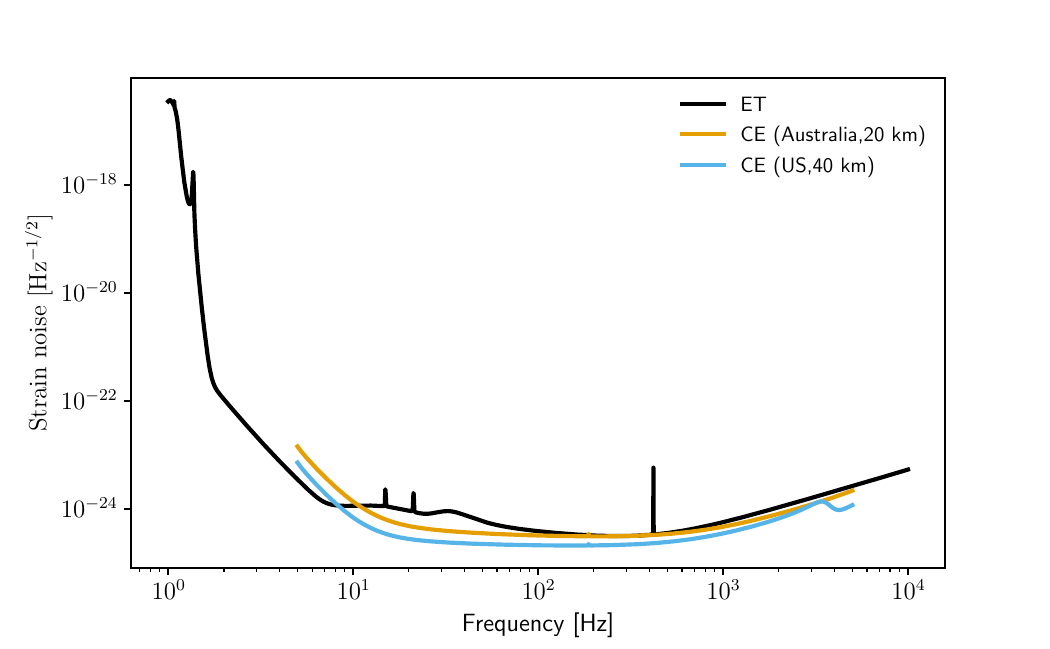}
\centering
\caption{\label{fig3} The sensitivity curves of ET and CE used in this work. Note that we consider one 20-km CE in Australia and one 40-km CE in the US.}
\end{figure}
%\begin{table*}[!htb]
\begin{table}
\setlength\tabcolsep{1mm}
\renewcommand{\arraystretch}{1}
\centering
\caption{\label{tab1}The Coordinates of the 3G GW Observatories Used in This Work \citep{di2021seismological,Borhanian:2020ypi}.}

\begin{tabular}{ccccc}
\hline\hline    & $\varphi$ & $\lambda$ & $\gamma$ & $\zeta$ \\
GW Detector  & $(\mathrm{deg})$ & $(\mathrm{deg})$ & $(\mathrm{deg})$ & $(\mathrm{deg})$ \\
\hline
% LIGO Hanford &  $46.455$ & $-119.408$ & $170.999$ & 90 \\
% LIGO Livingston &  $30.563$ & $-90.774$ & $242.716$ & 90 \\
% Virgo &  $43.631$ & $10.504$ & $115.567$ & 90 \\
% KAGRA &  $36.412$ & $137.306$ & $15.396$ & 90 \\
% LIGO India &  $19.613$ & $77.031$ & $287.384$ & 90 \\
Cosmic Explorer, US &  $43.827$ & $-112.825$ & $45.000$ & 90.000 \\
Einstein Telescope, Europe &  $40.443$ & $9.457$ & $0.000$ & 60.000 \\
Cosmic Explorer, Australia &  $-34.000$ & $145.000$ & $90.000$ & 90.000 \\
\hline
\end{tabular}
\begin{threeparttable}
\begin{tablenotes}[flushleft]
      \item \textbf{Note.} Here $\varphi$ is the latitude, $\lambda$ is the longitude, $\gamma$ is the orientation angle of the detector's arms measured counterclockwise from east of the Earth to the bisector of the interferometer arms, and $\zeta$ is the opening angle between the detector's two arms.
    \end{tablenotes}
\end{threeparttable}
\end{table}

\subsection{Mock Gravitational-wave Dark Siren Data}
\begin{figure*}[!htp]
\includegraphics[width=0.45\textwidth]{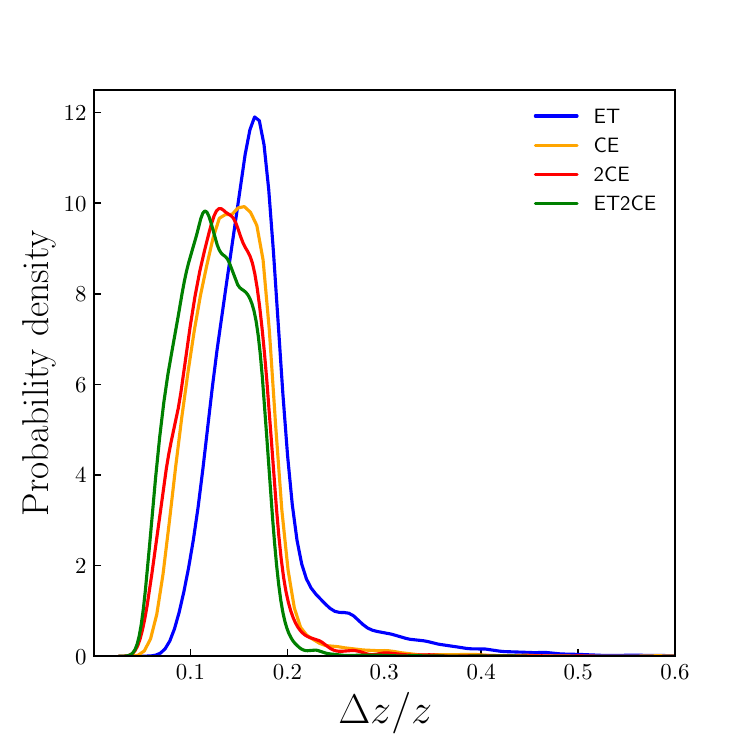} \ \hspace{1cm}
\includegraphics[width=0.45\textwidth]{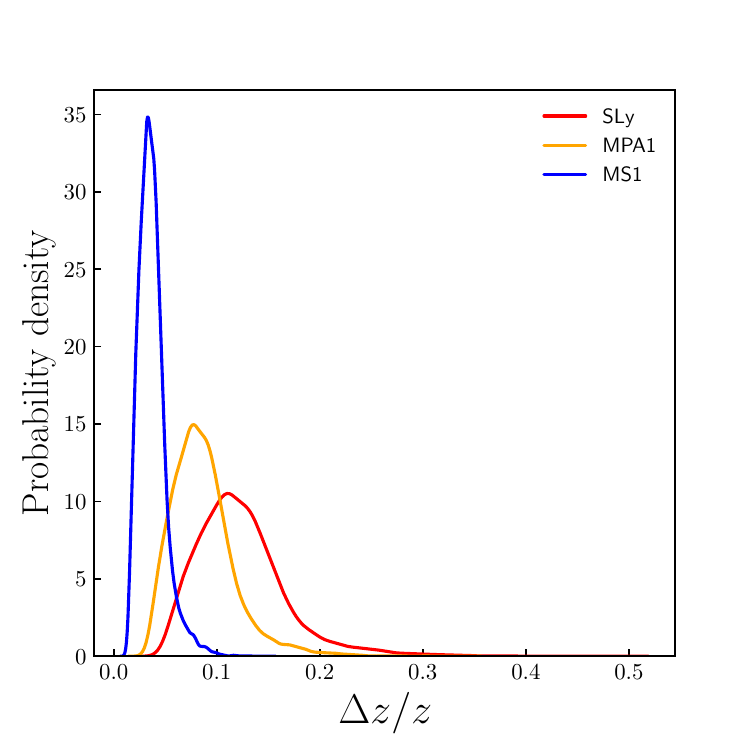}\ \hspace{1cm}
\centering \caption{\label{fig4} Relative errors of redshift distributions. Left panel: distributions of $\Delta z/z$ for ET, CE, 2CE, and ET2CE. Right panel: distributions of $\Delta z/z$ for SLy, MPA1, and MS1 based on the observation of ET2CE.}
\end{figure*}

Under the stationary phase approximation \citep{Zhang:2017srh}, the Fourier transform of the time-domain waveform for a detector network is given by \citet{Wen:2010cr,Zhao:2017cbb}
\begin{equation}
\tilde{\boldsymbol{h}}(f)=e^{-i \Phi} \hat{\boldsymbol{h}}(f),\label{boldsymbolh}
\end{equation}
where $\Phi$ is the $N\times N$ diagonal matrix with $\Phi_{ij}=2\pi f\delta_{ij}(\textbf{n}\cdot\textbf{r}_{a})$, $\textbf{n}$ is the propagation direction of a GW, $\textbf{r}_a$ is the location of the $a$th detector, and $\boldsymbol{h}$ is given by
\begin{equation}
\hat{\boldsymbol{h}}(f)=\Big[\tilde{h}_1 (f),\tilde{h}_2 (f),\cdots,\tilde{h}_N (f)\Big].
\end{equation}

In this work, we consider the waveform in the inspiralling stage for the nonspinning BNS system. We adopt the restricted post-Newtonian (PN) approximation and calculate the waveform to the 3.5 PN order \citep{Cutler:1992tc,Blanchet:2004bb,Sathyaprakash:2009xs}. The Fourier transform of the GW waveform of the $a$th detector is given by \citet{Sathyaprakash:2009xs}
\begin{align}
\tilde{h}_{a}(f)=\mathcal{A}_a&f^{-7/6}\exp \{i\big(2 \pi f t_{\rm c}-\pi / 4 +2 \psi(f / 2)\nonumber\\-&\varphi_{(2,0), a}+\Phi_{\text {tidal}}(f)\big)\},
\end{align}
where the Fourier amplitude $\mathcal{A}_a$ is given by
\begin{align}
\mathcal{A}_a=&~~\frac{1}{d_{\rm L}}\sqrt{\big(F_{+, a}(1+\cos^2\iota)\big)^2+(2F_{\times, a}\cos\iota)^2}\nonumber\\
            &~~\times \sqrt{5\pi/96}\pi^{-7/6}\mathcal{M}_{\rm chirp}^{5/6},
\end{align}
$\psi(f)$ and $\varphi_{(2,0), a}$ are given by \citet{Blanchet:2004bb,Sathyaprakash:2009xs}
\begin{align}
&\psi(f)=-\psi_{\rm c}+\frac{3}{256 \eta} \sum_{i=0}^{7} \psi_{i}(2\pi Mf)^{(i-5) / 3}, \\
&\varphi_{(2,0), a}=\tan ^{-1}\left(-\frac{2 \cos \iota F_{\times, a}}{\big(1+\cos ^{2}\iota\big) F_{+, a}}\right),
\end{align}
where $d_{\rm L}$ is the luminosity distance to the GW source, $F_{+,a}$ and $F_{\times,a}$ are the antenna response functions\footnote{The detailed forms of $F_{+,\times}$ can be found in e.g. \citet{Jaranowski:1998qm}, \citet{Arnaud:2001my}, and \citet{Schutz:2011tw}.} of the $a$th GW detector which are related to the locations of the GW source and the GW detector (the adopted coordinates of GW observatories are listed in Table~\ref{tab1} and shown in Figure~\ref{fig1}), $\iota$ is the inclination angle between the binary's orbital angular momentum and the line of sight, $\mathcal{M}_{\rm chirp}=(1+z)\eta^{3/5}M$ is the observed chirp mass, $M=m_1+m_2$ is the total mass of binary system with component masses $m_1$ and $m_2$, $\eta=m_1 m_2/(m_1+m_2)^2$ is the symmetric mass ratio, $\psi_{\rm c}$ is the coalescence phase, and the detailed forms of coefficients $\psi_{i}$ could be found in e.g., \citet{Sathyaprakash:2009xs}. Note that the GW waveform in the frequency domain is adopted, i.e., time $t$ is replaced by \citet{Maggiore:2007ulw}
\begin{align}
t_{f}=t_{\rm c}-(5 / 256) \mathcal{M}_{\rm chirp}^{-5 / 3}(\pi f)^{-8 / 3},
\end{align}
where $t_{\rm c}$ is the coalescence time.

The tidal contribution to the GW phase in the frequency domain is given by \citet{Messenger:2011gi}
\begin{align}
\Phi&_{\text {tidal }}(f)= \sum_{a=1}^{2} \frac{3 \lambda_{a}(1+z)^{5}}{128 \eta M^{5}}\bigg[-\frac{24}{\chi_{a}}\left(1+\frac{11 \eta}{\chi_{a}}\right)(\pi M f)^{5 / 3}\nonumber\\ & -\frac{5}{28 \chi_{a}}\left(3179-919 \chi_{a}-2286 \chi_{a}^{2}+260 \chi_{a}^{3}\right)(\pi M f)^{7 /3}\bigg],
\end{align}
where the sum is comprised of the components of the binary system, $\chi_a=m_a/(m_1+m_2)$, and $\lambda$ is the tidal deformation parameter, which is the function of the mass of an NS. Given that various EoSs of an NS can yield distinct redshift measurements, a stiffer EoS is deemed to provide more accurate redshift determinations \citep{Messenger:2011gi}.  
In order to investigate the impact of an NS's EoS on the cosmological parameter estimations, we chose the SLy \citep{Douchin:2001sv}, MPA1 \citep{Muther:1987xaa}, and MS1 \citep{Mueller:1996pm} models as representative EoSs of an NS, representing the soft, intermediate, and stiff EoSs of an NS, respectively.
We use the linear function to fit the $\lambda-m$ relation in the mass range of $[1.2, 2]\ M_{\odot}$, which is consistent with the current observations. We obtain $\lambda=Bm+C$ with $B=-2.03$ and $C=4.43$, $B=-1.37$ and $C=4.67$, and $B=-3.67$ and $C=12.69$ for the fitting of SLy, MPA1, and MS1.

For the BNS merger rate, we used the methodology detailed in \citet{Jin:2022qnj}. We assume the Madau-Dickinson star formation rate \citep{Madau:2014bja}, with an exponential time delay (an e-fold time of 100 Myr; \citealt{Vitale:2018yhm}). Meanwhile, we adopt the local comoving merger rate to be 320 $\rm Gpc^{-3}~yr^{-1}$, which is the estimated median rate based on the O3a observation \citep{LIGOScientific:2020kqk}. In Figure~\ref{fig2}, we show the expected detection rates as a function of redshift for ET, CE, the 2CE network, and the ET2CE network. Note that in the following discussions, we consider a 3 yr observation of the four detection strategies.

Throughout this work, we simulate the BNS mergers with random binary orientations and sky directions. The sky direction, binary inclination, polarization angle, and coalescence phase are randomly chosen in the ranges of ${\rm cos}\,\delta\in [-1,1]$, $\alpha\in [0,360^\circ]$, ${\rm cos}\,\iota\in [-1,1]$, $\psi\in [0,360^\circ]$, and $\psi_{\rm c}\in [0,360^\circ]$. For the mass of an NS, we adopt the mass distribution given by the latest LIGO-VIRGO-KAGRA observation \citep{LIGOScientific:2021aug}.

After simulating the GW catalog, we need to calculate the detection number of GW events. The signal-to-noise ratio (S/N) is given by
\begin{equation}
\rho^2=\langle\tilde{\boldsymbol{h}}| \tilde{\boldsymbol{h}}\rangle.
\end{equation}
The inner product is defined as
\begin{equation}
\left\langle \zeta| \xi\right\rangle=2 \int_{f_{\text {lower }}}^{f_{\mathrm{upper}}} \frac{{\zeta}(f) {\xi}^{*}(f)+{\zeta}^{*}(f) {\xi}(f)}{S_{\rm n}(f)} \mathrm{~d} f,
\end{equation}
where $*$ represents complex conjugate, $f_{\text {lower}}$ is the lower cutoff frequency, $f_{\text {upper}}=2 /\left(6^{3 / 2} 2 \pi M_{\text {obs}}\right)$ is the frequency at the last stable orbit with $M_{\mathrm{obs}}=\left(m_{1}+m_{2}\right)(1+z)$, and $S_{\rm n}(f)$ is the one-side power spectral density (PSD). The adopted PSD forms for ET \footnote{https://www.et-gw.eu/index.php/etsensitivities/} and CE \footnote{https://cosmicexplorer.org/sensitivity.html} are shown in Figure~\ref{fig3}. Note that we take into account the triangular configuration of ET in the calculation of S/N.
We adopt the threshold of S/N to be 8 in the simulation.

\begin{table*}[!htb]
\caption{\label{tab2}The Absolute Errors (1$\sigma$) and the Relative Errors of the Cosmological Parameters.}
\centering
\setlength\tabcolsep{2pt}
\renewcommand\arraystretch{2}
\begin{tabular}{cccccccccccc}
\hline\hline
          Model       & Error &ET & CE & 2CE & ET2CE & CMB & CMB+ET & CMB+CE & CMB+2CE & CMB+ET2CE & CMB+BAO+SN    \\ \hline
 \multirow{3}{*}{I$\Lambda$CDM1}& $\sigma(\beta)$& $0.0490$ & $0.0390$   & $0.0310$ & $0.0230$ & $0.4800$ & $0.0210$   & $0.0160$ & $0.0130$ & $0.0094$ & $0.1000$ \\
 &$\sigma(H_0)$ &$0.180$ &$0.130$ & $0.100$ & $0.080$ &$4.600$ &$0.160$ & $0.110$ & $0.074$ & $0.053$ & $0.820$ \\
  &$\sigma(\Omega_{\rm m})$&$0.0110$ &$0.0078$  & $0.0057$ & $0.0050$ &$0.1500$ &$0.0087$  & $0.0063$ & $0.0049$ & $0.0035$ & $0.0290$ \\ \hline
 \multirow{3}{*}{I$\Lambda$CDM2}& $\sigma(\beta)$& $0.0450$ & $0.0210$  & $0.0160$ & $0.0130$ & $0.0023$ & $0.0008$  & $0.0007$ & $0.0007$ & $0.0007$ & $0.0012$ \\
 &$\sigma(H_0)$ &$0.180$ &$0.110$ & $0.083$ & $0.071$ &$1.700$ &$0.093$ & $0.076$ & $0.049$ & $0.041$ & $0.650$ \\
  &$\sigma(\Omega_{\rm m})$&$0.0077$ &$0.0043$  & $0.0033$ & $0.0027$ &$0.0250$ &$0.0019$  & $0.0016$ & $0.0010$ & $0.0009$ & $0.0081$ \\ \hline
\multirow{3}{*}{I$\Lambda$CDM3}& $\sigma(\beta$)& $0.0560$ & $0.0430$   & $0.0360$ & $0.0240$ & $0.530$ & $0.0310$   & $0.0270$ & $0.0220$ & $0.0180$ & $0.1500$ \\
 &$\sigma(H_0)$ &$0.190$ &$0.130$ & $0.097$ & $0.081$ &$3.300$ &$0.150$ & $0.110$ & $0.081$ & $0.063$ & $0.820$ \\
  &$\sigma(\Omega_{\rm m})$&$0.0120$ &$0.0083$  & $0.0065$ & $0.0054$ &$0.1300$ &$0.0091$  & $0.0074$ & $0.0057$ & $0.0049$ & $0.0350$ \\ \hline
 \multirow{3}{*}{I$\Lambda$CDM4}& $\sigma(\beta)$& $0.0460$ & $0.0390$   & $0.0280$ & $0.0180$ & $0.2200$ & $0.0250$   & $0.0220$ & $0.0190$ & $0.0170$ & $0.0440$ \\
 &$\sigma(H_0)$ &$0.170$ &$0.120$ & $0.098$ & $0.075$ &$4.800$ &$0.140$ & $0.120$ & $0.085$ & $0.073$ & $0.810$ \\
  &$\sigma(\Omega_{\rm m})$&$0.0079$ &$0.0056$  & $0.0043$ & $0.0030$ &$0.1500$ &$0.0050$  & $0.0044$ & $0.0034$ & $0.0030$ & $0.0160$ \\ \hline
\end{tabular}
\begin{threeparttable}
\begin{tablenotes}[flushleft]
      \item \textbf{Note.} Models considered: I$\Lambda$CDM1 ($Q=\beta H\rho_{\rm de}$), I$\Lambda$CDM2 ($Q=\beta H\rho_{\rm c}$), I$\Lambda$CDM3 ($Q=\beta H_0\rho_{\rm de}$), and I$\Lambda$CDM4 ($Q=\beta H_0\rho_{\rm c}$). Data used: ET, CE, 2CE, ET2CE, CMB, CMB+ET, CMB+CE, CMB+2CE, CMB+ET2CE, and CMB+BAO+SN. Here, $H_0$ is in units of $\rm {km~s^{-1}~Mpc^{-1}}$.
    \end{tablenotes}
\end{threeparttable}
\end{table*}

Then we use the Fisher information matrix to estimate the instrumental error of the luminosity distance. The Fisher information matrix of a network including $N$ independent GW detectors is given by
\begin{equation}
\boldsymbol{F}_{ij}=\left\langle\frac{\partial \tilde{\boldsymbol{h}}}{\partial \theta_i}\Bigg|  \frac{\partial \tilde{\boldsymbol{h}}}{\partial \theta_j}\right\rangle,
\end{equation}
where $\boldsymbol{\theta}$ denotes 12 parameters ($d_{\rm L}$, $z$, $\mathcal{M}_{\rm chirp}$, $\eta$, $t_{\rm c}$, $\psi_{\rm c}$, $\iota$, $\delta$, $\alpha$, $\psi$, $B$, $C$) for a given BNS system. The covariance matrix is equal to the inverse of the Fisher information matrix, and thus we can use the Fisher information matrix to calculate the measurement errors of the above 12 GW parameters.
The error of the parameter $\theta_i$ can be calculated by
\begin{equation}
\Delta \theta_{i}=\sqrt{\left(F^{-1}\right)_{i i}},\label{eq:error}
\end{equation}
where $F$ is the total Fisher information matrix.

For the measurement errors of $d_{\rm L}$, we consider the instrumental error $\sigma_{d_{\mathrm{L}}}^{\mathrm{inst}}$ and weak-lensing error $\sigma_{d_{\mathrm{L}}}^{\mathrm{lens}}$. The total error of $d_{\rm L}$ is
\begin{equation}
\left(\sigma_{d_{\mathrm{L}}}\right)^{2}=\left(\sigma_{d_{\mathrm{L}}}^{\mathrm{inst}}\right)^{2}+\left(\sigma_{d_{\mathrm{L}}}^{\text {lens }}\right)^{2}. \label{total}
\end{equation}
The instrumental error of $d_{\rm L}$ could be obtained by Eq.~(\ref{eq:error}), i.e., $\sigma_{\rm d_{\rm L}}^{\rm inst}=\Delta d_{\rm L}=\sqrt{(F^{-1})_{11}}$.
The error caused by weak lensing is adopted from \citet{Hirata:2010ba}, \citet{Tamanini:2016zlh}, \citet{Speri:2020hwc},
\begin{align}
\sigma_{d_{\rm L}}^{\rm lens}(z)=&~~\left[1-\frac{0.3}{\pi / 2} \arctan \left(z / 0.073\right)\right]\times d_{\rm L}(z)\nonumber\\
        &~~\times
        0.066\bigg[\frac{1-(1+z)^{-0.25}}{0.25}\bigg]^{1.8}.
\end{align}

For the measurement error of redshift $\Delta z$ using the dark siren method, i.e., $\Delta z=\sqrt{(F^{-1})_{22}}$, we show the relative errors of redshift distributions for ET, CE, 2CE, and ET2CE in the left panel of Figure~\ref{fig4}. We see that the measurement precisions of redshifts are mainly 10\%-30\%, which is consistent with the result given in \citet{Jin:2022qnj}

\subsection{Cosmological Data}
For comparison, we also employ the current cosmological data, i.e., the CMB data, the baryon acoustic oscillation (BAO) data, and the type Ia supernova (SN) data. For the CMB data, we employ the Planck TT, TE, EE spectra at $\ell\geq 30$, the low-$\ell$ temperature Commander likelihood, and the low-$\ell$ SimAll EE likelihood from the Planck 2018 data release \citep{Planck:2018vyg}. For the BAO data, we adopt the measurements from 6dFGS ($z_{\rm eff}=0.106$;  \citealt{Beutler:2011hx}), SDSS-MGS ($z_{\rm eff}=0.15$; \citealt{Ross:2014qpa}), and BOSS DR12 ($z_{\rm eff}=0.38$, 0.51, and 0.61; \citealt{BOSS:2016wmc}). For the SN data, we use the Pantheon+ sample, which contains 1550 samples in the range 0.001 $< z <$ 2.26 \citep{Brout:2022vxf}. We adopt the Markov Chain Monte Carlo method \citep{Lewis:2002ah} to maximize the likelihood and infer the posterior distributions of cosmological parameters. The detailed calculation of the likelihood can be referred to in \citet{Jin:2022qnj}.

\section{Results and Discussions}\label{sec3}
\begin{figure*}[!htp]
\includegraphics[width=0.45\textwidth]{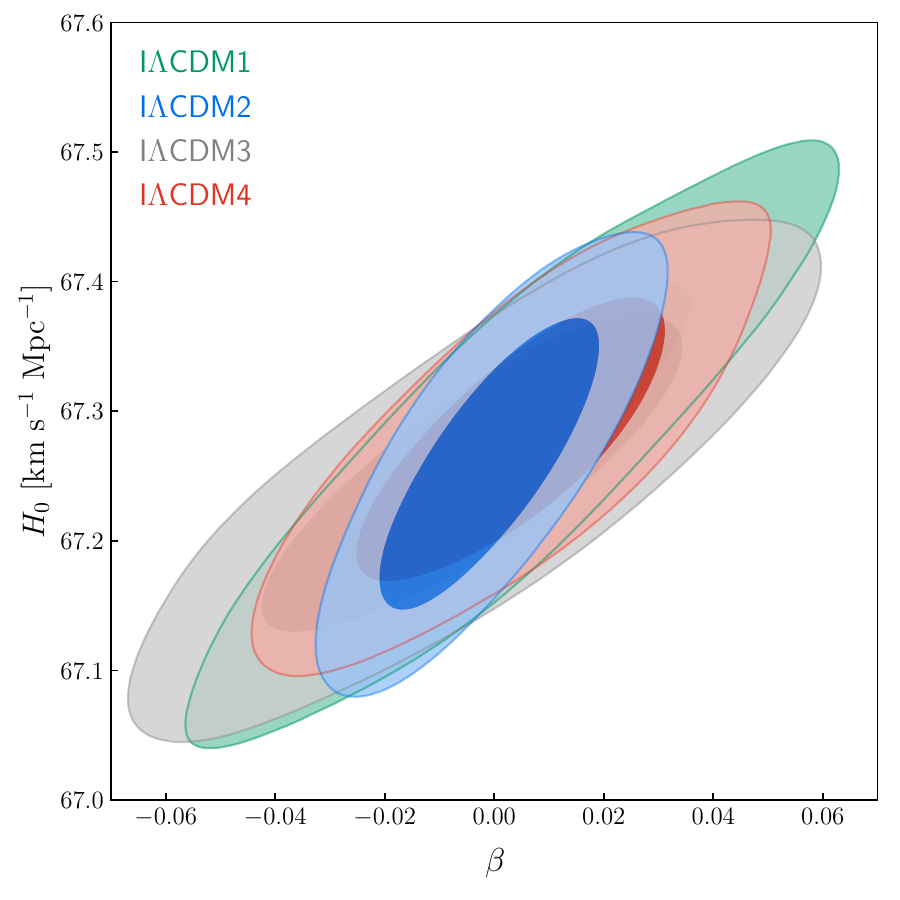} \ \hspace{1cm}
\includegraphics[width=0.45\textwidth]{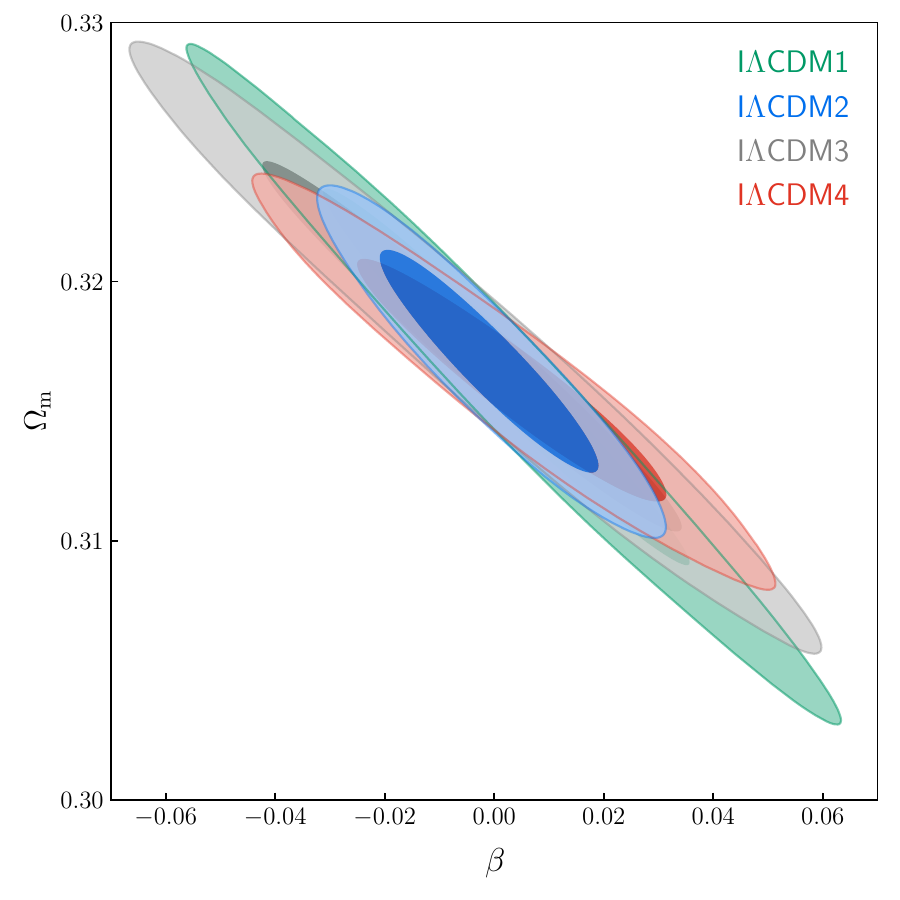}\ \hspace{1cm}
\centering \caption{\label{fig5} Two-dimensional marginalized contours (68.3\% and 95.4\% confidence level) in the $\beta$--$H_0$ and $\beta$--$\Omega_{\rm m}$ planes by using the ET2CE data in the I$\Lambda$CDM1, I$\Lambda$CDM2, I$\Lambda$CDM3, and I$\Lambda$CDM4 models.}
\end{figure*}

\begin{figure*}[!htp]
\includegraphics[width=0.45\textwidth]{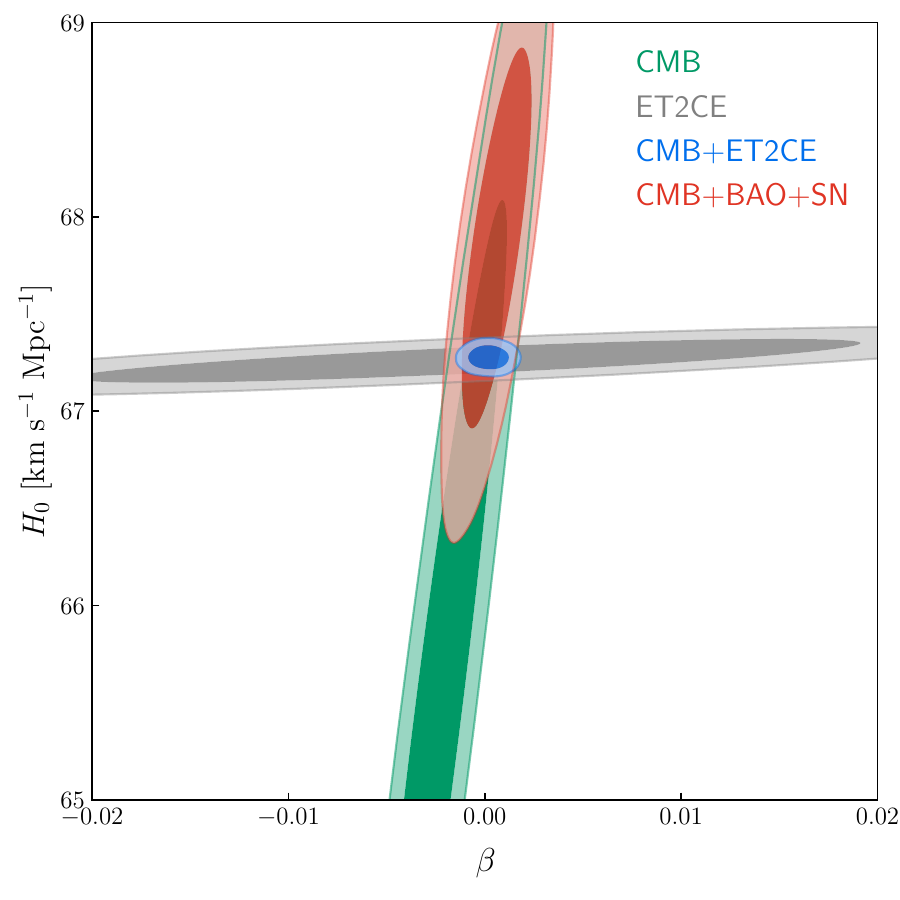} \ \hspace{1cm}
\includegraphics[width=0.45\textwidth]{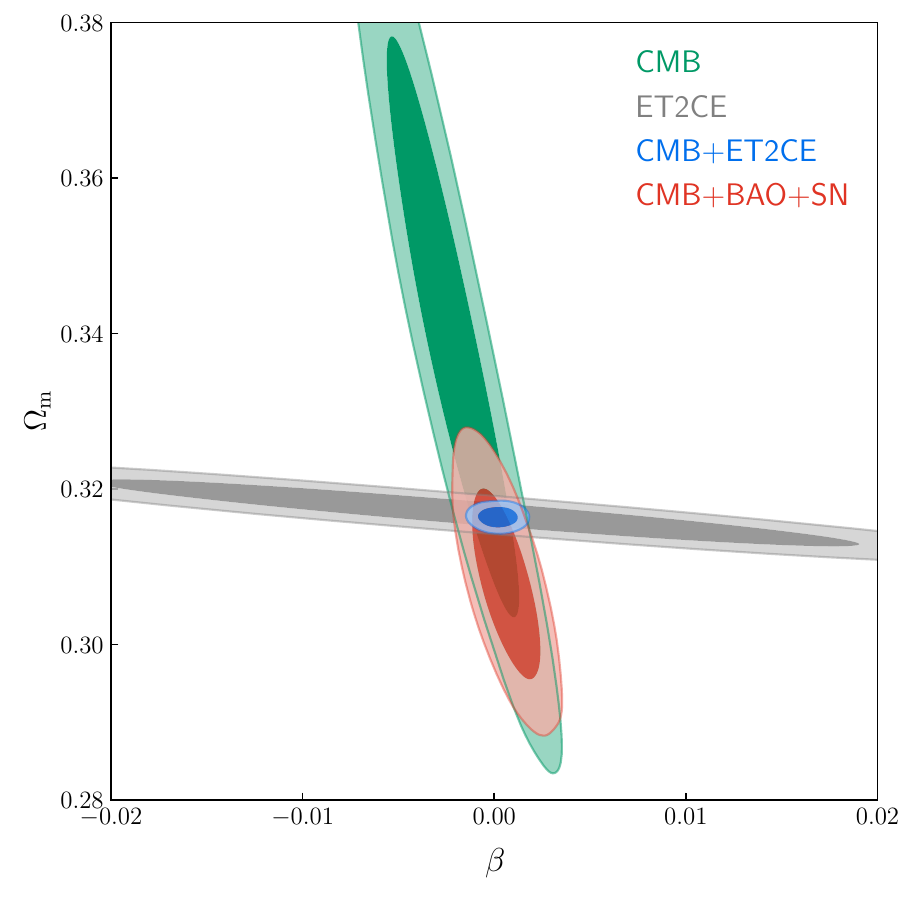}\ \hspace{1cm}
\centering \caption{\label{fig6} {Two-dimensional marginalized contours (68.3\% and 95.4\% confidence level) in the $\beta$--$H_0$ and $\beta$--$\Omega_{\rm m}$ planes by using the CMB, ET2CE, CMB+ET2CE, and CMB+BAO+SN data for the I$\Lambda$CDM2 model.}}
\end{figure*}

\begin{figure*}[!htp]
\includegraphics[width=0.45\textwidth]{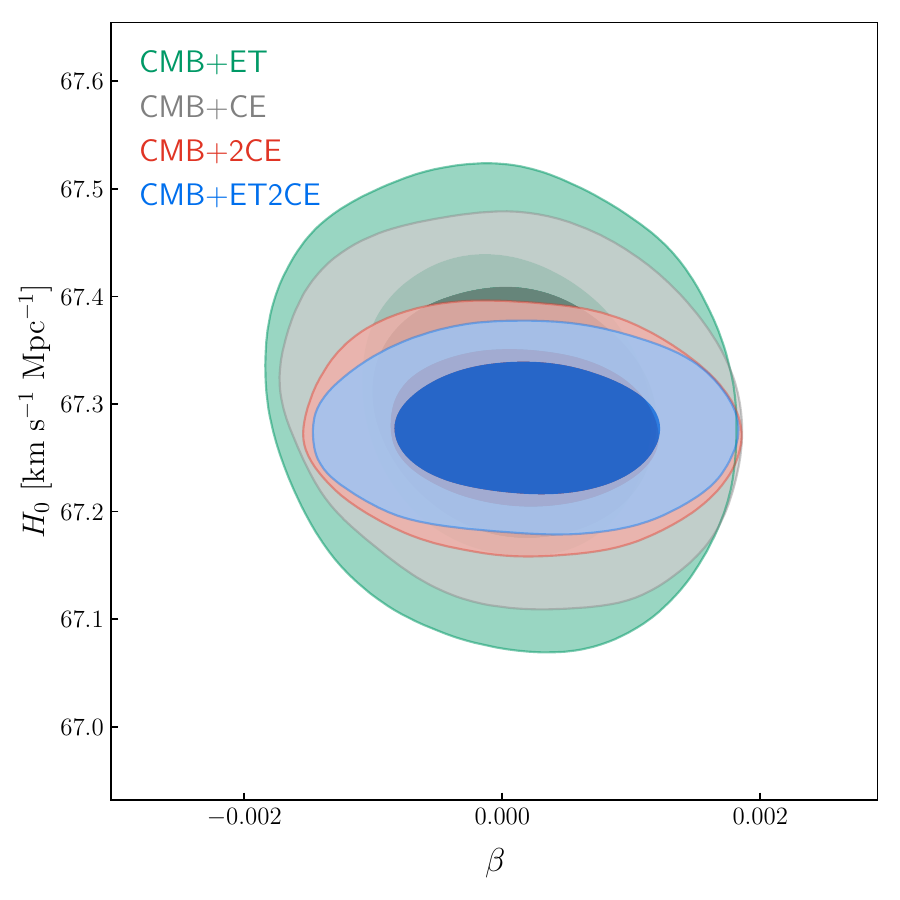} \ \hspace{1cm}
\includegraphics[width=0.45\textwidth]{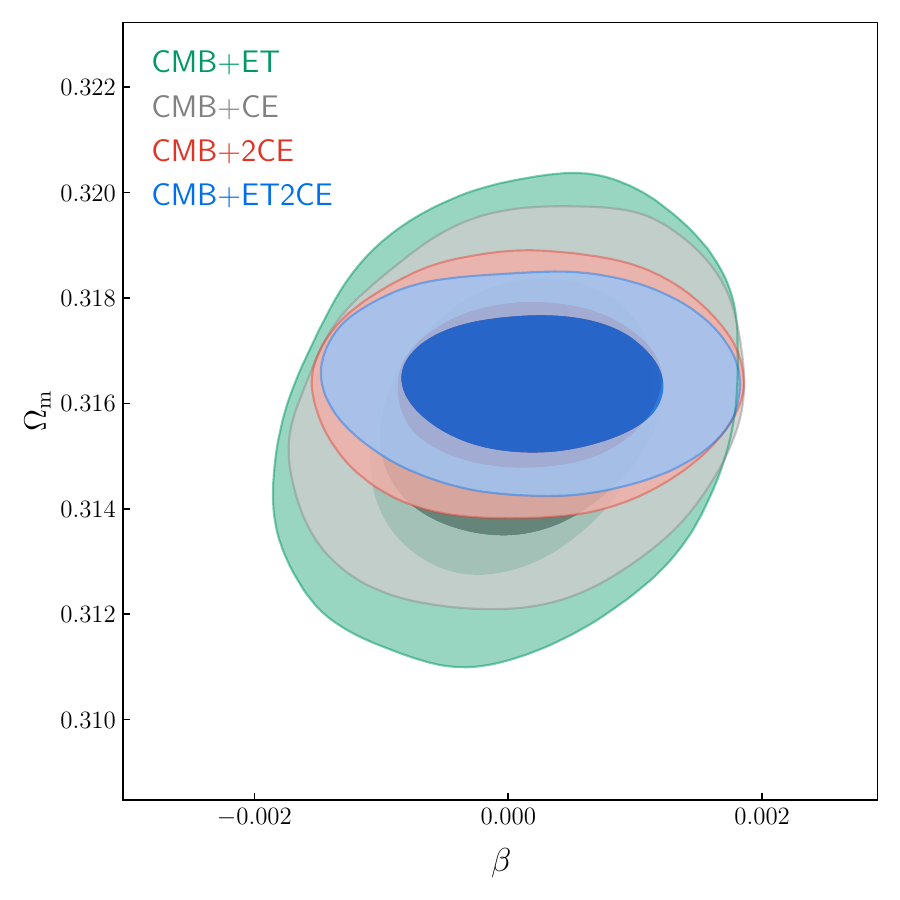}\ \hspace{1cm}
\centering \caption{\label{fig7} {Two-dimensional marginalized contours (68.3\% and 95.4\% confidence level) in the $\beta$--$H_0$ and $\beta$--$\Omega_{\rm m}$ planes by using the CMB+ET, CMB+CE, CMB+2CE, and CMB+ET2CE data for the I$\Lambda$CDM2 model.}}
\end{figure*}

\begin{figure*}[!htp]
\includegraphics[width=0.45\textwidth]{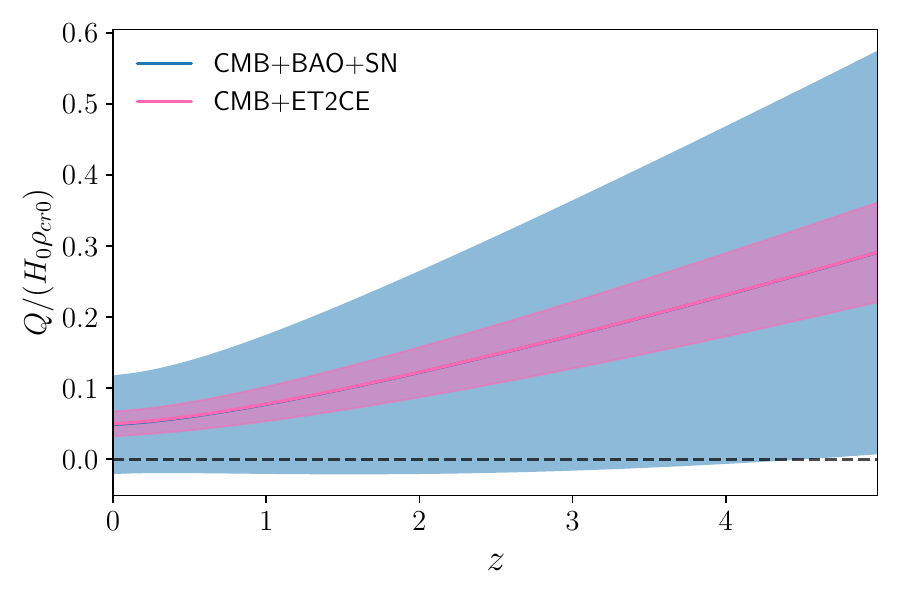} \ \hspace{1cm}
\includegraphics[width=0.45\textwidth]{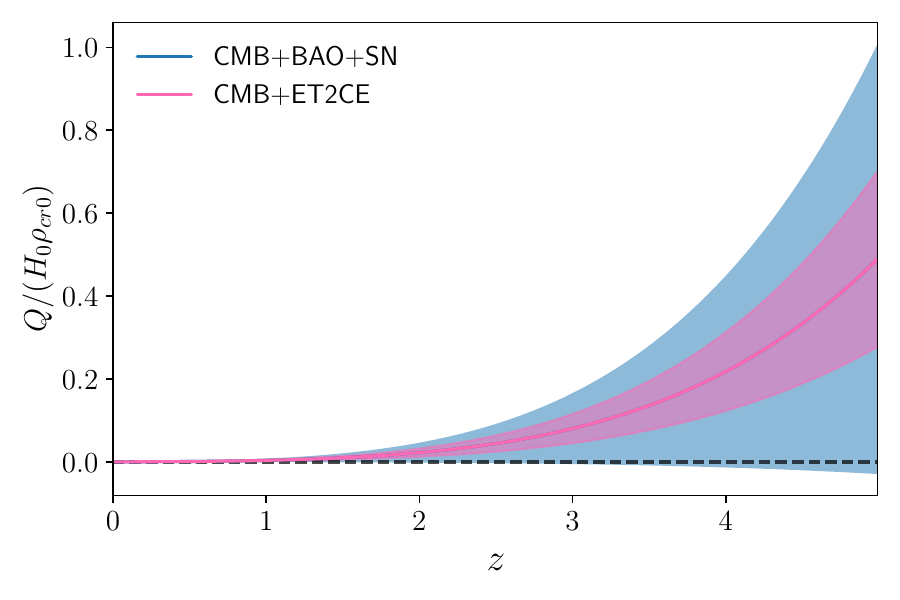}\ \hspace{1cm}
\centering \caption{\label{fig8} {The evolutions of $Q/(H_0 \rho_{\rm cr0})$ in the I$\Lambda$CDM1 (left panel)
and I$\Lambda$CDM2 (right panel) models. The blue and pink shaded regions represent the $1\sigma$ constraints from the CMB+BAO+SN and CMB+ET2CE data, respectively. The black dashed lines denote the $\Lambda$CDM model ($Q=0$).}}
\end{figure*}

\begin{figure*}[!htp]
\includegraphics[width=0.45\textwidth]{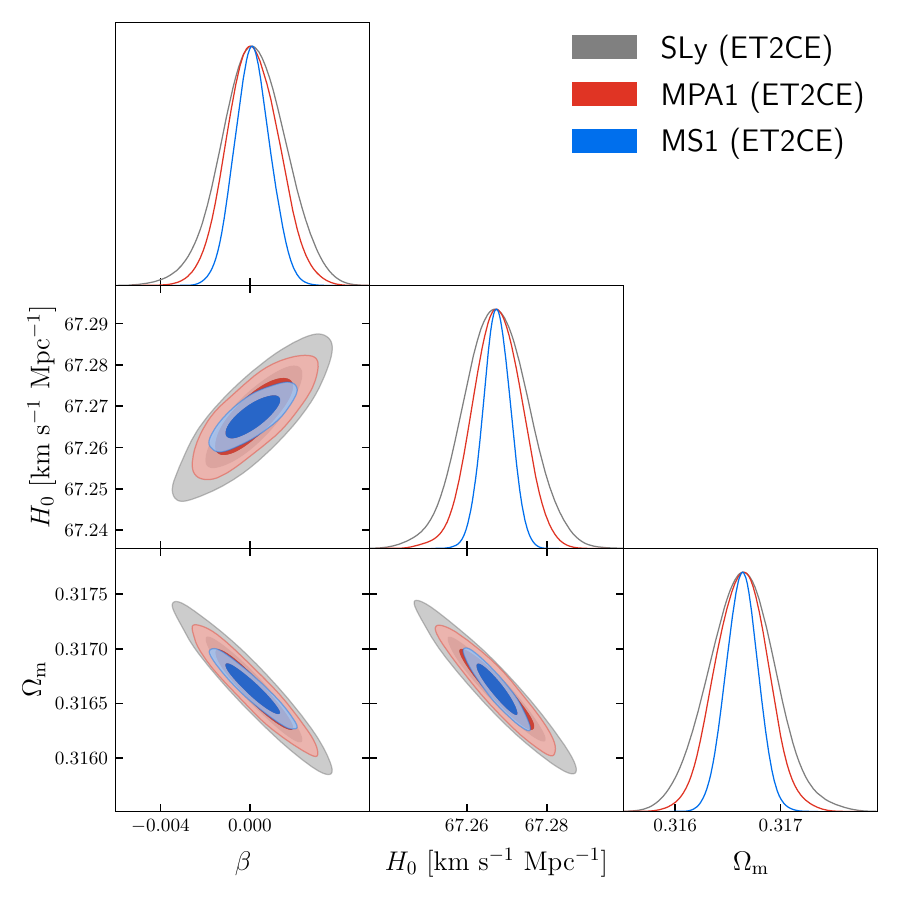} \ \hspace{1cm}
\includegraphics[width=0.45\textwidth]{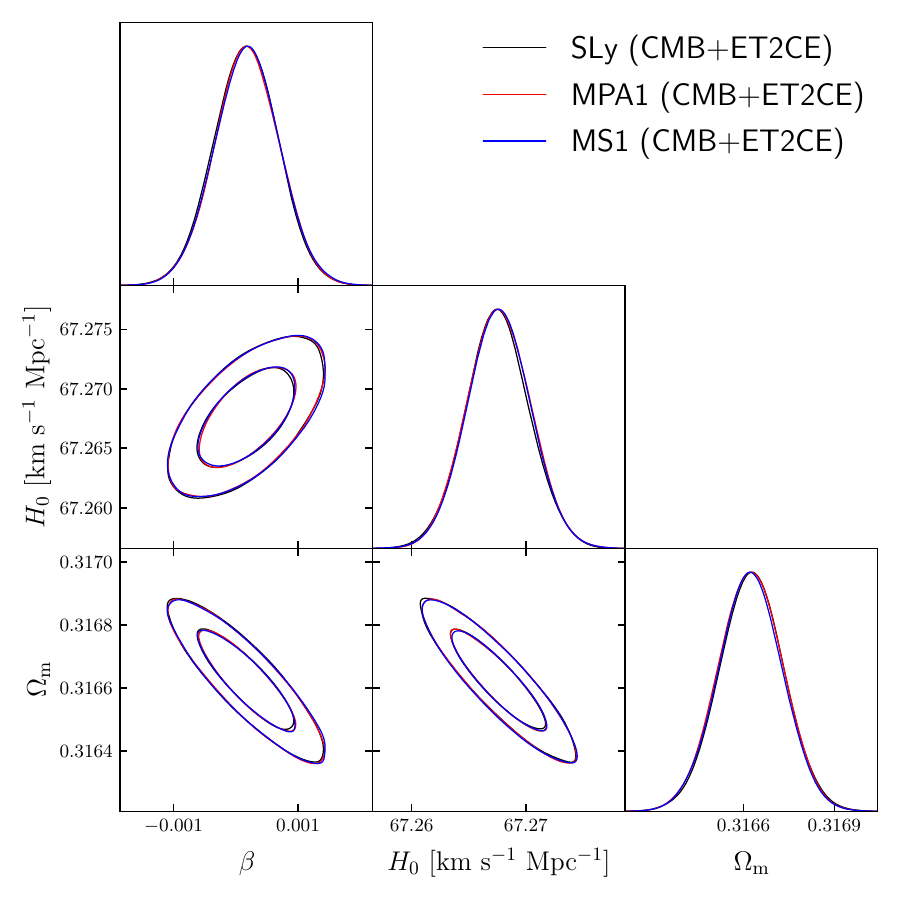}\ \hspace{1cm}
\centering \caption{\label{fig9} {Constraints on cosmological parameters using dark sirens from ET2CE based on the SLy, MPA1, and MS1 models. Left panel: constraints on the cosmological parameters using the dark sirens from ET2CE based on the SLy, MPA1, and MS1 models in the I$\Lambda$CDM2 model. Right panel: constraints on the cosmological parameters using the  CMB+ET2CE data based on the SLy, MPA1, and MS1 models in the I$\Lambda$CDM2 model.}}
\end{figure*}

\begin{table}
\caption{\label{tab3}The 1$\sigma$ Errors of the Cosmological Parameters in the I$\Lambda$CDM2 Model.}
\centering
\setlength{\tabcolsep}{3mm}{
\renewcommand{\arraystretch}{1.5}
%\begin{tabular}{l|cccc}
\begin{tabular}{p{3cm}cccc}
%\bottomrule[1pt]
\hline\hline
              Data      & $\sigma(\beta)$ & $\sigma(H_0)$ & $\sigma(\Omega_{\rm m})$  \\
\hline
ET2CE (SLy)              & 0.0130 & 0.071                             & 0.0027             \\
ET2CE (MPA1)          & 0.0110 & 0.065                             & 0.0012                \\
ET2CE (MS1)             & 0.0080 & 0.031                            & 0.0005               \\
CMB+ET2CE (SLy)             & 0.0007 & 0.041                             & 0.0009               \\
CMB+ET2CE (MPA1)      & 0.0007 & 0.041                           & 0.0009             \\
CMB+ET2CE (MS1)      & 0.0007 & 0.040                           & 0.0009             \\
\hline
%Fiducial value     & 0       & 67.36                            & 0.3153           & 0.8111     \\
%\bottomrule[1pt]
\end{tabular}}
\begin{threeparttable}
\begin{tablenotes}[flushleft]
      \item \textbf{Note.} Data used: ET2CE (SLy), ET2CE (MPA1), ET2CE (MS1), CMB+ET2CE (SLy), CMB+ET2CE (MPA1), and CMB+ET2CE (MS1). Here, $H_0$ is in units of $\rm {km~s^{-1}~Mpc^{-1}}$.
    \end{tablenotes}
\end{threeparttable}
\end{table}

In this section, we shall report the constraint results of the cosmological parameters. We consider the I$\Lambda$CDM1 ($Q=\beta H\rho_{\rm de}$), I$\Lambda$CDM2 ($Q=\beta H\rho_{\rm c}$), I$\Lambda$CDM3 ($Q=\beta H_0\rho_{\rm de}$), and I$\Lambda$CDM4 ($Q=\beta H_0\rho_{\rm c}$) models to perform cosmological analysis. For the GW observations, we consider four detection strategies, i.e., ET, CE, the 2CE network, and the ET2CE network, to constrain the cosmological models. In order to show the ability of GW dark sirens to break cosmological parameter degeneracies, we also show the constraint results of CMB, CMB+ET, CMB+CE, CMB+2CE, CMB+ET2CE, 
and CMB+BAO+SN. The posterior distribution contours ($1\sigma$ and 2$\sigma$) for the interested cosmological parameters are shown in Figures~\ref{fig5}--\ref{fig7} and summarized in Table~\ref{tab2}. The evolutions of $Q/(H_0 \rho_{\rm cr0})$ in the I$\Lambda$CDM1 and I$\Lambda$CDM2 models are shown in Figure~\ref{fig8}. In order to show the impact of different EoSs of an NS on estimating cosmological parameters using these dark sirens, we choose three typical EoSs of an NS to perform cosmological analysis and show the constraint results in Figure~\ref{fig9}.

\subsection{Constraints on the Interacting Dark Energy models}
In Figure~\ref{fig5}, we show the constraint results of the ET2CE network in the $\beta$--$H_0$ (left panel) and $\beta$--$\Omega_{\rm m}$ (right panel) planes for the I$\Lambda$CDM1, I$\Lambda$CDM2, I$\Lambda$CDM3, and I$\Lambda$CDM4 models. From the left panel, we can see that I$\Lambda$CDM2 gives the best constraints on cosmological parameters, followed by I$\Lambda$CDM4, while I$\Lambda$CDM1 and I$\Lambda$CDM3 give comparable constraint results. Concretely, in the case of I$\Lambda$CDM2, ET2CE gives $\sigma(\beta)=0.0130$ and $\sigma(H_0)=0.071~\rm km~s^{-1}~Mpc^{-1}$, which is 43.5\% and 11.3\% better than those of I$\Lambda$CDM1. From the right panel, we can see that the above results still hold. As shown in Table~\ref{tab2}, while ET2CE gives the best constraint results among ET, CE, 2CE, and ET2CE, it still does not performs well in constraining the dimensionless coupling parameter $\beta$, whereas GW dark sirens can all precisely measure the Hubble constant. Since I$\Lambda$CDM2 gives the best constraint results among the IDE models, we choose I$\Lambda$CDM2 as a representative to make the following discussion.

In Figure~\ref{fig6}, in order to show the ability of GW dark sirens to break the cosmological parameter degeneracies, we show the constraints in the $\beta$--$H_0$ and $\beta$--$\Omega_{\rm m}$ planes for the I$\Lambda$CDM2 model. We can see that the parameter degeneracy orientations of CMB and ET2CE are almost orthogonal, and thus, their combination can effectively break the cosmological parameter degeneracies. Furthermore, the contours of CMB+ET2CE are much smaller than those of CMB+BAO+SN, indicating that the ability of dark sirens from ET2CE to break the cosmological parameters is much better than that of the current BAO and SN data. Concretely, CMB+ET2CE gives $\sigma(\beta)=0.00068$, $\sigma(H_0)=0.041~\rm km~s^{-1}~Mpc^{-1}$, and $\sigma(\Omega_{\rm m})=0.00087$, which are 70.4\%, 97.6\%, and 96.5\% better than those of CMB, respectively. Moreover, the constraints on $\beta$, $H_0$, and $\Omega_{\rm m}$ using CMB+ET2CE are 43.3\%, 93.7\%, and 89.3\% are better than those using CMB+BAO+SN, respectively.

In Figure~\ref{fig7}, we show the constraint results of CMB+ET, CMB+CE, CMB+2CE, and CMB+ET2CE for the I$\Lambda$CDM2 model. We can see that CMB+ET2CE also gives the best constraint results, followed by CMB+2CE, CMB+CE, and CMB+ET, but the difference is not as big as the constraint results of the single GW dark sirens, as shown in Table~\ref{tab2}. Moreover, we can see that even CMB+ET gives much better constraint results than the current CMB+BAO+SN data, showing the potential of the GW dark sirens in breaking the cosmological parameter degeneracies.

In Figure~\ref{fig8}, we show the evolutions of $Q/(H_0 \rho_{\rm cr0})$ versus $z$ in the I$\Lambda$CDM1 and I$\Lambda$CDM2 models. Here $\rho_{\rm cr0}$ is the present-day critical density of the universe. The best-fit line and $1\sigma$ regions are shown in the figure. In the I$\Lambda$CDM1 model, CMB+ET2CE gives much better constraint than the CMB+BAO+SN in the whole redshift region, indicating that the dark sirens from ET2CE can provide great help in constraining the evolution of $Q$. While in the I$\Lambda$CDM2 model, CMB+BAO+SN can provide a precise constraint on $\beta$, whereas CMB+ET2CE can provide a much tighter constraint on the evolution of $Q$.

\subsection{Impact of the Equation of State of Neutron Stars on Performing Cosmological Analysis}

In this subsection, we wish to investigate the impact of an NS's EoS on estimating cosmological parameters. As pointed out in \citet{Messenger:2011gi}, different EoSs of an NS could lead to different measurement precision of redshift when measuring the redshift from the tidal deformation of an NS. Hence, we wish to offer an answer to the question of what impact an NS's EoS will have on estimating the cosmological parameters. Here we choose three typical (soft, intermediate, and stiff) EoSs of an NS, i.e., SLy, MPA1, and MS1, to perform cosmological analysis. Note that the following discussions are based on the I$\Lambda$CDM2 model using ET2CE. See Figure~\ref{fig9} and Table~\ref{tab3} for the results of the cosmologicalconstraints.

In the left panel of Figure~\ref{fig9}, we show the constraint results of ET2CE based on SLy, MPA1, and MS1. We find that the stiffer MS1 gives tight constraints than the softer MPA1 and SLy. This is because the stiffer EoS can lead to more precise redshift measurement, as shown in the right panel of Figure~\ref{fig4}, which is also consistent with the redshift measurement forecasted in \citet{Messenger:2011gi}. As mentioned above, the GW dark sirens can effectively break the cosmological parameter degeneracies generated by the EM observation. In the right panel of Figure~\ref{fig9}, we show the constraint results of the combination of CMB and GW dark sirens. We find that the combination of CMB and GW dark sirens shows basically the same constraint results. This means that in the IDE scenario, an NS's EoS does have an impact on estimating cosmological parameters, whereas the combination results of CMB and GW dark siren are basically not affected by the EoS of an NS.

\section{Conclusion}\label{sec4}
In this work, we investigate the potential of the GW dark sirens from the 3G GW detectors in probing the interaction between dark energy and dark matter. Based on the 3 yr observation, we consider four detection strategies, i.e., ET, CE, the 2CE network, and the ET2CE network, to perform cosmological analysis. Four phenomenological IDE models are considered, i.e., I$\Lambda$CDM1 ($Q=\beta H\rho_{\rm de}$), I$\Lambda$CDM2 ($Q=\beta H\rho_{\rm c}$), I$\Lambda$CDM3 ($Q=\beta H_0\rho_{\rm de}$), and I$\Lambda$CDM4 ($Q=\beta H_0\rho_{\rm c}$).

We find that GW dark sirens can provide tight constraints on $\Omega_{\rm m}$ and $H_0$ in all the IDE models but do not perform well in constraining $\beta$ in the I$\Lambda$CDM2 model. In the other IDE models, the constraints on $\beta$ are all better than the result of the current CMB+BAO+SN data. Nonetheless, we find that in the I$\Lambda$CDM2 model, the parameter degeneracy orientations of CMB and GW dark sirens are almost orthogonal, and thus, the combination of them could effectively break the cosmological parameter degeneracies and significantly improve the measurement precisions of the cosmological parameters. With the addition of the GW dark sirens to the CMB data, the constraints on cosmological parameters could be improved by 70.4\%--97.6\%. CMB+ET2CE gives $\sigma(\beta)=0.00068$ in the I$\Lambda$CDM2 model, which is much better than that of the current CMB+BAO+SN data.

In addition, we investigate the impact of the EoS of an NS on estimating cosmological parameters. The stiffer EoS could give tighter constraints than the softer EoS. However, the combination of CMB and GW dark sirens based on different EoSs of an NS shows basically the same constraint results, implying that the EoS of an NS has almost no impact on estimating cosmological parameters in the case of the combination of CMB and GW dark sirens. Our results show that the GW dark sirens from the 3G GW detectors are expected to play an important role in helping probe the interaction between dark energy and dark matter.

\section*{acknowledgments}

We thank Rui-Qi Zhu, Ji-Yu Song, and Ji-Guo Zhang for their helpful discussions. This work was supported by the National SKA Program of China (grants Nos. 2022SKA0110200 and 2022SKA0110203) and the National Natural Science Foundation of China (grants Nos. 11975072, 11875102, 11835009, and 12305068).

\bibliography{ide_tidal}{}
\bibliographystyle{aasjournal}

\end{document}